\documentclass{appolb}

\usepackage{epsfig}% Include figure files

\begin{document}
%\eqsec
\pagestyle{plain}
%\newcount\eLine\eLine=\inputlineno\advance\eLine by -1

%\title{The asymmetry term in the nuclear-matter incompressibility from measurements on the giant monopole resonance: An Update
\title{THE ASYMMETRY TERM IN THE NUCLEAR-MATTER INCOMPRESSIBILITY FROM MEASUREMENTS ON THE GIANT MONOPOLE RESONANCE: AN UPDATE
\thanks{Presented at the 2010 Zakopane Conference on Nuclear Physics}
}

\author{U. Garg
\address{Physics Department, University of Notre Dame, Notre Dame, IN 46556, USA}
}
\maketitle

\begin{abstract}

We have investigated the isoscalar giant monopole resonances (ISGMR) in $^{112-124}$Sn and $^{106-116}$Cd nuclei using inelastic scattering of 386-MeV $\alpha$-particles at extremely forward angles, including 0$^\circ$.
%We have obtained completely ``background-free'' inelastic-scattering spectra %over the angular range $0^\circ$-$9^\circ$.
The strength distributions for various multipoles were extracted by a multipole decomposition analysis based on the expected angular distributions of the respective multipoles. From the ISGMR results, a value of $K_{\tau} \sim -500$ MeV is obtained for the asymmetry term in the nuclear incompressibility.

\end{abstract}


\section{Introduction}
Investigation of giant resonances has had a rich history over the past six decades since the discovery of the isovector giant dipole resonance (IVGDR)~\cite{ivgdr1,ivgdr2,ivgdr3}. Of the various giant resonances that have been identified and investigated since, the compression-mode giant resonances--the isoscalar giant monopole resonance (ISGMR; the ``breathing mode'') and the isoscalar giant dipole resonance (ISGDR; the ``squeezing mode'')--occupy the pride of place because their energy is directly related to the nuclear incompressibility, $K_{A}$, from which the incompressibility of infinite nuclear matter, $K_{\infty}$, may be deduced \cite{blaizot}. The latter is critical in our understanding of a number of interesting phenomena from collective excitations of nuclei to supernova explosions and radii of neutron stars.

In this review, we discuss some recent results on the ISGMR over a series of isotopes of Sn and Cd. This data has provided an ``experimental'' value for the asymmetry term, $K_{\tau}$, of nuclear incompressibility; this term, associated with the neutron excess $(N - Z)$ is important, for example, in obtaining radii of neutron stars in EOS calculations.

\section{Experimental Techniques}
The experiment was performed at the ring cyclotron facility of the Research Center for Nuclear Physics (RCNP), Osaka University, using inelastic scattering of 386-MeV $\alpha$ particles at extremely forward angles, including 0$^\circ$. Details of the experimental procedures and data-analysis techniques have been provided elsewhere \cite{li1,li2} and are only briefly described here. Self-supporting target foils of enriched $^{112, 114, 116, 118, 120, 122, 124}$Sn and $^{106, 110, 112, 114, 116}$Cd isotopes of thickness $\sim$5--10 mg/cm$^2$ were employed; special target frames with a large aperture were used in order to reduce the background caused by the beam-halo hitting the frames. Data were also taken with a $^{nat}$C target at the actual field settings used in the experiments and energy calibration was obtained from the peak positions of the 7.652- and 9.641-MeV states in the $^{12}$C$(\alpha,\alpha')$ spectra.

Inelastically-scattered $\alpha$ particles were momentum-analyzed with the high-resolution magnetic spectrometer ``Grand Raiden''~\cite{fuji} and the vertical and horizontal positions of the $\alpha$ particles were measured with a focal-plane detector system comprised of two position-sensitive multi-wire drift chambers (MWDCs) and two scintillators~\cite{itoh}. The MWDCs and scintillators enabled us to make particle identification and to reconstruct the trajectories of the scattered particles. The scattering angle at the target and the momentum of the scattered particles were determined by the ray-tracing method. The vertical-position spectrum obtained in the double-focusing mode of the spectrometer was exploited to eliminate the instrumental background~\cite{itoh,uchida}. Examples of such ``free-from-instrumental-background'' inelastic scattering spectra for the Sn isotopes have been provided in Refs. \cite{li1,li2}.

A multipole decomposition analysis (MDA) procedure~\cite{bonin} was employed to extract the strengths of the ISGMR, along with those for other isoscalar resonances (up to $L$ = 3). The associated DWBA calculations were performed with the computer code PTOLEMY~\cite{ptolemy}, following the method of Satchler and Khoa \cite{satchler}, using the density-dependent single-folding model for the real part, obtained with a Gaussian $\alpha$-nucleon potential, and a phenomenological Woods-Saxon potential for the imaginary term. The optical-model (OM) parameters were determined by fitting the differential cross sections of elastic $\alpha$ scattering measured in a companion experiment; the efficacy of the OM parameters was tested by comparing the experimental cross sections for the first 2$^+$ states in these nuclei with calculated cross sections using a collective form factor and previously-established $B(E2)$ values. Again, examples of MDA fits to the experimental angular distributions of the differential cross sections have been provided in Refs. \cite{li1,li2}.

\section{Results and Discussion}
\begin{figure}[htb]
\includegraphics[width=12.4 cm]{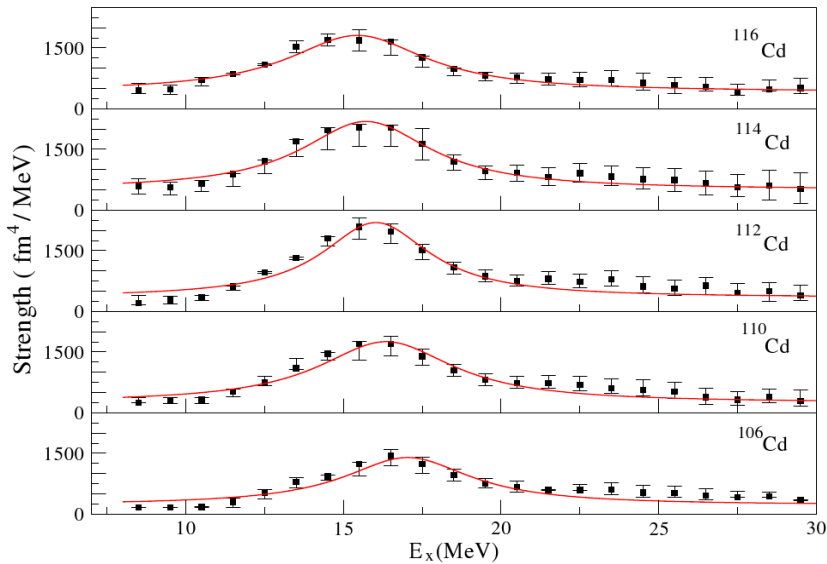}
\caption{\label{fig1} ISGMR strength distributions obtained for the Cd isotopes in the present experiment. Error bars represent the uncertainties from fitting the angular distributions in the MDA procedure.
The solid lines show Lorentzian fits to the data. {\bf These results should be considered preliminary}.}
\end{figure}

We have extracted strength distributions for $L$=0, 1, 2, and 3 multipoles over the energy range 8.5 MeV--31.5 MeV in all the Sn and Cd isotopes investigated in this work. The ISGMR strength distributions for the Cd isotopes are presented in Fig. 1. The $L=0$ strength distributions were fitted with a Lorentzian function to determine the centroid energies and widths of the ISGMR. These fits are shown superimposed in Fig.~\ref{fig1}; the corresponding fitting parameters are presented in Table~1. Also presented are the various moment ratios for the experimental ISGMR strength distributions calculated over the excitation-energy range, $E_x$ = 10.5--20.5 MeV, encompassing the ISGMR peak. The results for the Sn isotopes have been presented previously~\cite{li1,li2}.

A note of caution is apt here. There is a small, near-constant ISGMR strength up to the highest excitation energies measured in this experiment. The exact reason behind this extra strength is not quite well understood. However, similarly enhanced E1 strengths at high excitation energies were noted previously \cite{itoh,uchida2} and have been attributed to contributions to the continuum from three-body channels, such as knockout reactions \cite{brand}. These processes are implicitly included in the MDA as background and may lead to spurious contributions to the extracted multipole strengths at higher energies where the associated cross sections are very small.

\begin{table}[!h]
\caption{\label{table1}Lorentzian-fit parameters, as extracted from MDA and the various moment ratios for the ISGMR strength distributions in the Cd isotopes. All moments have been calculated over $E_x$ = 10.5--20.5 MeV. {\bf These results should be considered preliminary}.}

\begin{tabular}{lccccc}
\hline 
\multicolumn{1}{c}{Target} & E$_{GMR} (MeV)$&$\Gamma (MeV)$ & $\frac{m_1}{m_0}$ (MeV)&$\sqrt{\frac{m_1}{m_{-1}}}$ (MeV)&$\sqrt{\frac{m_3}{m_1}}$ (MeV)\\
\hline \\

$^{106}$Cd & $16.4\pm0.1$ & $4.4\pm0.5$ & $16.7\pm0.1$ & $16.5\pm0.2$ & $17.2\pm0.3$\\
$^{110}$Cd & $16.1\pm0.1$ & $4.1\pm0.6$ & $16.3\pm0.1$ & $16.1\pm0.2$ & $16.9\pm0.3$\\
$^{112}$Cd & $15.8\pm0.1$ & $4.9\pm0.7$ & $16.2\pm0.1$ & $16.0\pm0.2$ & $16.8\pm0.2$\\
$^{114}$Cd & $15.5\pm0.1$ & $5.0\pm0.6$ & $16.1\pm0.1$ & $15.8\pm0.2$ & $16.7\pm0.4$\\
$^{116}$Cd & $15.4\pm0.1$ & $5.0\pm0.4$ & $15.9\pm0.1$ & $15.7\pm0.1$ & $16.6\pm0.3$\\

\hline \\
\end{tabular}
\end{table}

%As mentioned earlier, the primary focus of this work has been on the ISGMR because of its direct connection %with the nuclear incompressibility.
The excitation energy of the ISGMR is expressed in the scaling model~\cite{stringari} as:
\begin{equation}
\label{eqn:gmr} E_{ISGMR}=\hbar\sqrt{\frac{K_A}{m<r^2>}}
\end{equation}

\noindent where $m$ is the nucleon mass, $<r^2>$ the ground-state
mean-square radius, and $K_{A}$, the incompressibility of the
nucleus.

Further, the incompressibility of a nucleus, $K_{A}$, may be expressed as:
\begin{equation}
K_{A} \sim  K_{vol}(1 + cA^{-1/3}) + K_{\tau}((N - Z)/A)^{2} +
K_{Coul}Z^{2}A^{-4/3}
\end{equation}

\noindent Here, $c \approx$ -1 (see, for example, Ref. \cite{skp04}, and $K_{Coul}$ is essentially model independent (in the sense that the deviations from one theoretical model to another are quite small), so that the associated term can be calculated for a given isotope. Thus, for a series of isotopes, the difference $K_A-K_{Coul}Z^2A^{-4/3}$ may be approximated to have a quadratic relationship with the asymmetry parameter ((N - Z)/A)), of the type y = A + Bx$^2$, with $K_\tau$ being the coefficient, B, of the quadratic term. It has been established previously~\cite{shlomo2,pears} that direct fits to the Eq.~2 do not provide good constraints on the value of $K_\infty$. However, this expression has been used in this case not to obtain a value for $K_\infty$, but only to demonstrate the approximately quadratic relationship between $K_A$ and the asymmetry parameter. In addition, it should be understood that this expression provides only an ``average'' value for $K_\tau$ and the available data is not sensitive to higher-order effects like the ``surface'' part of this term.
\noindent Here, $c \approx$ -1 (see, for example, Ref. \cite{skp04}, and $K_{Coul}$ is essentially model independent (in the sense that the deviations from one theoretical model to another are quite small), so that the associated term can be calculated for a given isotope. Thus, for a series of isotopes, the difference $K_A-K_{Coul}Z^2A^{-4/3}$ may be approximated to have a quadratic relationship with the asymmetry parameter ((N - Z)/A)), of the type y = A + Bx$^2$, with $K_\tau$ being the coefficient, B, of the quadratic term. It has been established previously~\cite{shlomo2,pears} that direct fits to the Eq.~2 do not provide good constraints on the value of $K_\infty$. However, this expression has been used in this case not to obtain a value for $K_\infty$, but only to demonstrate the approximately quadratic relationship between $K_A$ and the asymmetry parameter. In addition, it should be understood that this expression provides only an ``average'' value for $K_\tau$ and the available data is not sensitive to higher-order effects like the ``surface'' part of this term.

From such an analysis of the ISGMR data in the Sn isotopes, we had obtained a value of $K_\tau = -550\pm100$~MeV (see Fig. 4 in Ref.~\cite{li1}). An identical analysis of the ISGMR data in the Cd isotopes gives a preliminary value of $K_\tau = -480\pm100$~MeV (see Fig. 2). These numbers are in good agreement with each other, and are also consistent with the value of $K_{\tau}~=~-370\pm120$~MeV obtained from an analysis of the isotopic transport ratios in medium-energy heavy-ion reactions \cite{bao3}, the value $K_{\tau}=-500^{+125}_{-100}$~MeV obtained by  Centelles {\em et al.} \cite{spain} from constraints put by neutron-skin data from anti-protonic atoms across the mass table, and $K_{\tau}=-500\pm50$ MeV obtained by
Sagawa {\em et al.} by comparing our Sn ISGMR data with calculations using different Skyrme Hamiltonians and RMF Lagrangians \cite{sagawa7}.
A more precise determination of K$_\tau$ will likely result from extending the ISGMR measurements to longer isotopic chains. This provides strong motivation for measuring the ISGMR strength in unstable nuclei, a focus of current investigations at the new rare isotope beam facilities at RIKEN, GANIL, GSI, and NSCL.

\begin{figure}[htb]
\includegraphics[width=12.4cm]{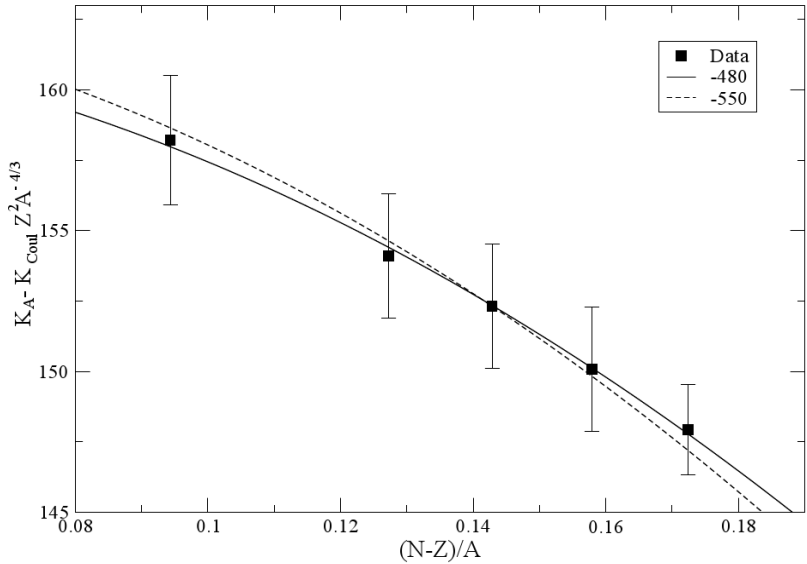}
\caption{\label{fig2} Systematics of the difference
$K_A - K_{Coul}Z^{2}A^{-4/3}$ in the Cd isotopes as a function of the
``asymmetry-parameter'' ((N-Z)/A); $K_{Coul}$ = -5.2$\pm$0.7 MeV~\cite{sagawa7}. The solid line represents a
least-square quadratic fit to the data. The dashed line shows, for comparison, results with the value $K_\tau$ = -550 MeV previously obtained from the Sn data. {\bf These results should be considered preliminary}.}
\end{figure}

From the ISGMR and ISGDR data, one now has a consistent value for $K_\infty = 240\pm20$~MeV \cite{garg4,colo,jorge,shlomo}.\footnote{We are using a more conservative uncertainty in the value than that cited previously.} Combined with the ``experimental'' value of $K_\tau$, this can provide a means of selecting the most appropriate of the interactions used in EOS calculations. For example, this combination of ``experimental'' values for $K_\infty$ and $K_\tau$ appears to rule out a vast majority of the Skyrme-type interactions currently in use in nuclear structure calculations~\cite{sagawa6,ugbulk}.

Pearson {\em et al.}~\cite{pearson} have recently suggested that our data for the Sn isotopes may be fitted with 
$K_\tau$ = -300 MeV, a value they claim is more amenable to many theoretical calculations. They fix the $K_\tau $ 
term at this value and by introducing a higher-order term, $K_{ss} A^{1/3}$, with $K_\tau$ in the expansion, obtain a reasonable 
fit to the data. In effect, they have changed the fitting expression from y = A + Bx$^2$ (which we had used) to 
y = A' + (B'+C')x$^2$ and fixed A' and B', to obtain a value for C' from fit to the data. Clearly, there are several 
problems with their approach. First, considering that C' would change very little ($\sim$3\%) over the range of 
the fit while the x$^2$ term changes by ~230\%, C' is, for all practical purposes (and definitely for the uncertainties 
involved in the extraction of the coefficient), a constant. So, the fit of Pearson {\em et al.} replaces one constant 
with two and, of course, one can find a large number of combinations of B' and C' that would give essentially the 
same fit as the one given by B! [Because Pearson {\em et al.} choose to fix B' at -300 MeV, they obtain a value 
of -1900$A^{1/3}$ MeV for C'. Indeed, if they had chosen the value of -100 MeV for B', they would get the same 
fit for C' = -2880$A^{1/3}$ MeV!]. Second, they do not discuss at all the ``reasonableness'' of the value for 
$K_{ss}$ (-1900 MeV) that is obtained in their fit.  Nominally, one expects the higher order terms to be smaller 
than the leading terms. In this case, the higher-order term is nearly 30\% larger! In this context, we note that the 
value for $K_{ss}$ that they obtain with $K_\tau$ = -550 MeV, leads to the higher-order term being only about 
30\% of the leading-order term, a much more reasonable situation. Finally, there is also the issue of their use of 
a value of 370 MeV for the surface term $K_{sf}$ (=$cK_{vol}$ in Eq. 2 above). Per almost all theoretical estimates, 
the value of $c \simeq$ 1 (see Refs. \cite{skp04,sagawa7}). Pearson {\em et al.} obtain $c$ = 1.54 but, again, do 
not justify a number at such variance from other theoretical work.
\section{Summary}
We have measured the strength distributions of the isoscalar giant monopole
resonance (ISGMR) in the even-A
Sn and Cd isotopes via inelastic scattering of 386-MeV $\alpha$
particles at extremely forward angles, including 0$^{\circ}$.
The asymmetry-term, $K_{\tau}$, in the expression for the nuclear incompressibility has been determined to be $\sim$ -500 MeV from the ISGMR data and is found to be consistent with a number of indirectly extracted values for this parameter. Measurements with the new rare isotope beam facilities would go a long way in extending the isotopic chains, thus greatly reducing the uncertainty in the extracted value of $K_\tau$.

\section{Acknowledgments}
It is with profound gratitude that I acknowledge my collaborators in this work:
G.P.A. Berg, T.~Li, Y.~Liu, R. Marks, B.K.~Nayak, D. Patel, and P.V.~Madhusudhana Rao (University of Notre Dame); M.~Fujiwara, H.~Hashimoto, K.~Kawase, K.~Nakanishi, S.~Okumura, and M.~Yosoi (RCNP, Osaka, Japan); M.~Itoh, M.~Ichikawa, R.~Matsuo, T.~Terazano, and H.P. Yoshida (Tohuku University, Sendai, Japan); M.~Uchida (Tokyo Institute of Technology, Tokyo, Japan); H.~Akimune (Konan University, Kobe, Japan); Y.~Iwao, T.~Kawabata, T.~Murakami, H.~Sakaguchi, S.~Terashima, Y.~Yasuda and J.~Zenihiro (Kyoto University, Kyoto, Japan); and, M.N. Harakeh (KVI, Groningen, The Netherlands). Thanks are also due to the RCNP staff for providing high-quality $\alpha$ beams required for these measurements. This work has been supported in part by the US-Japan Cooperative Science Program of the JSPS, and by the U.S. National Science Foundation (Grants No. INT03-42942, PHY04-57120, and PHY07-58100).

\end{document}